\documentclass[aps,prb,twocolumn,superscriptaddress,nofootinbib,showpacs]{revtex4}

\usepackage{epsfig}
\usepackage{amsmath}
\usepackage{dcolumn}
\usepackage{multirow}

\newcommand{\Uppsala}{Division of High Energy Physics, Uppsala University. S-75121 Uppsala, Sweden}
\newcommand{\Irvine}{Department of Physics and Astronomy, University of
California, Irvine, CA 92697, USA}
\newcommand{\Bartol}{Bartol Research Institute, University of Delaware, Newark, DE 19716, USA}
\newcommand{\Wuppertal}{Fachbereich 8 Physik, BUGH Wuppertal, D-42097 Wuppertal, Germany}
\newcommand{\Brussels}{Universit\'e Libre de Bruxelles, Science Faculty CP230, B-1050 Brussels,~Belgium}
\newcommand{\DESY}{DESY-Zeuthen, D-15735 Zeuthen, Germany}
\newcommand{\Kalmar}{Department of Technology, Kalmar University, S-39182 Kalmar, Sweden}
\newcommand{\LBNL}{Lawrence Berkeley National Laboratory, Berkeley, CA 94720, USA}
\newcommand{\Berkeley}{Department of Physics, University of California, Berkeley, CA 94720, USA}
\newcommand{\Mainz}{Institute of Physics, University of Mainz, D-55099 Mainz, Germany}
\newcommand{\Philadelphia}{Department of Physics and Astronomy, University of Pennsylvania, Philadelphia, PA 19104, USA}
\newcommand{\Wisconsin}{Department of Physics, University of Wisconsin-Madison, WI 53706, USA}
\newcommand{\RF}{Department of Physics, University of Wisconsin-River Falls, WI 54022, USA}
\newcommand{\Stockholm}{Department of Physics, SCFAB, Stockholm University, S-10691 Stockholm, Sweden}
\newcommand{\Vrije}{Vrije Universiteit, Dienst ELEM, B-1050 Brussels,~Belgium}

\begin{document}

%\preprint{}

\title{Limits to the muon flux from WIMP annihilation in the center of the 
Earth with the AMANDA detector}

\author{J.~Ahrens}\affiliation{\Mainz}
\author{E. Andr\'es}\affiliation{\Stockholm}
\author{X.~Bai}\affiliation{\Bartol}
\author{G.~Barouch}\affiliation{\Wisconsin}
\author{S.W.~Barwick}\affiliation{\Irvine}
\author{R.C.~Bay}\affiliation{\Berkeley}
\author{T.~Becka}\affiliation{\Mainz}
\author{K.-H.~Becker}\affiliation{\Mainz}
\author{D.~Bertrand} \affiliation{\Brussels}
\author{A.~Biron}\affiliation{\DESY}
% ...

\collaboration{The AMANDA collaboration}
\noaffiliation

\date{\today}

\begin{abstract}
  A search for nearly vertical up-going muon-neutrinos from 
neutralino annihilations in the center of the Earth has been performed 
with the AMANDA-B10 neutrino detector. The data collected 
in 130.1 days of live-time in 1997, $\sim$10$^9$ events, have been
analyzed for this search. No excess over the expected atmospheric
neutrino background has been observed. An upper limit at 90\% confidence 
level has been obtained on the annihilation 
rate of neutralinos in the center of the Earth, as well as the
corresponding muon flux limit, both as a function of the
neutralino mass in the range 100~GeV-5000~GeV.

\end{abstract}

%\keywords{}

\maketitle

\section{\label{sec:Intro} Introduction}

 There are strong observational indications for the
existence of dark matter in the universe. 
Measurements of the energy density of the universe, 
$\Omega_{\mbox{\tiny 0}}$, from the combined analysis of 
cosmic microwave background radiation data and high red-shift Type Ia supernovae 
favor $\Omega_{\mbox{\tiny 0}}\,=\,1$, with a matter, 
$\Omega_{\mbox{\tiny M}}$, and a cosmological constant,  
$\Omega_{\tiny{\Lambda}}$, component. 
Combined with data from rotation curves of galaxies and cluster mass 
measurements, the matter contribution to $\Omega_{\mbox{\tiny 0}}$ is
$0.3\leq \Omega_{\mbox{\tiny M}}\leq 0.4$. 
Big Bang nucleosynthesis calculations of primordial 
helium, lithium and deuterium production, supported by abundance
measurements of these elements, set an upper limit 
on the amount of baryonic matter that can exist in the
universe, $\Omega_{\mbox{\tiny B}}\leq 0.05$ (see
Ref.~\onlinecite{Bergstrom:00a} for a recent review of values of 
$\Omega$). 
Non-baryonic dark matter must therefore constitute
a substantial fraction of $\Omega_{\mbox{\tiny M}}$.\par

 In this paper we present results of a search for non-baryonic dark matter in the 
form of weakly interacting massive particles (WIMP) using the 
AMANDA high-energy neutrino detector. The next section contains a brief motivation 
for WIMPs as dark matter candidates. Section~\ref{sec:detector}
describes briefly the characteristics of the AMANDA detector in the 
configuration used for this analysis. Sections~\ref{sec:simulations}
and~\ref{sec:analysis} contain a description of the simulation and analysis 
techniques used. In section~\ref{sec:systematics} we discuss the sources of the 
current systematic uncertainties of our analysis. In section~\ref{sec:results} 
we present the results of the analysis and we introduce a novel way of 
calculating  upper limits in the presence of systematic uncertainties. 
An upper limit on the neutrino-induced muon flux  
expected from WIMP annihilation in the center of the Earth is derived with 
this method.  A comparison with published muon-flux limits obtained by existing
neutrino experiments is presented in section~\ref{sec:comparison}.

\section{\label{sec:WIMPs}WIMPs as dark matter candidates}

 Particle physics provides an interesting dark matter candidate as a 
Weakly Interacting Massive Particle (WIMP). The relic density of  
particle type $i$ depends on its annihilation cross section, $\sigma$, as
$\Omega_i h^2\sim 3\times 10^{-27}/\left<\sigma {\mbox{v}}\right>$ 
(neglecting mass-dependent logarithmic corrections), where 
$\left<\right>$ indicates thermal average and v is the relative
velocity of the particles involved in the collision (see for example Ref.~\onlinecite{Jungman:96a}). 
Weak interactions provide the right annihilation cross section for the WIMPs to decouple 
in the early universe and give a relic density within the required range to contribute 
substantially to the energy density of the universe today. 
This is basically what would be needed to solve the dark matter problem. \par

In particular, and starting from a completely different rationale,
the Minimal Supersymmetric extension to the Standard Model of particle
physics (MSSM) provides a promising WIMP candidate in the neutralino,
$\chi$. The neutralino is a linear combination of the 
B-ino, $\tilde{\mbox{B}}$, and the W-ino, $\tilde{\mbox{W}}$,  
the supersymmetric partners of the electroweak gauge bosons, and of
the  H$^0_1$ and H$^0_2$, the neutral Higgs bosons, and 
it is stable (assuming R-parity conservation, which is further
supported to avoid too rapid proton decay). The actual composition of the 
neutralino can have cosmological consequences since its annihilation 
cross section depends on it. For example, it has been argued that a mainly 
W-ino type neutralino would not be cosmologically relevant in the
present epoch since it would have annihilated too fast in the early universe to 
leave any relevant relic density~\cite{Griest:90a}.

Still, the large parameter space of minimal supersymmetry
can be exploited to build realistic models which provide relic
neutralino densities within the 
cosmologically interesting region of 0.025$\lesssim \Omega_{\chi}h^2<$1.
Negative results from searches for supersymmetry at the LEP 
accelerator at CERN have set a lower  
limit on the neutralino mass  $m_{\chi}>$~31~GeV (Ref.~\onlinecite{OPAL:00a}), while 
theoretical arguments based on the requirement of unitarity set an upper limit of
340~TeV (Ref.~\onlinecite{Griest:90a}). Imposing in addition the 
condition on $\Omega_{\chi}h^2$ mentioned above, only models with 
m$_{\chi} \lesssim $ 10~TeV (Ref.~\onlinecite{Edsjo:97a}) become
cosmologically interesting.\par

 Neutralinos have a non-negligible probability of scattering off 
nuclei of ordinary matter.  Assuming the dark matter in the Galactic halo 
is (at least partially) composed of relic neutralinos,  
elastic interactions of these particles with  nuclei in the Earth can lead to 
energy losses that bring the neutralino below the escape velocity, 
 becoming gravitationally trapped~\cite{Press:85a,Freese:86a}. 
 For high neutralino masses ($>$ a few hundred GeV) direct capture 
from the halo population by the Earth is kinematically suppressed~\cite{Gould:88a}. In this 
case neutralinos can be accreted from the population already captured by the solar system. 
Gravitational capture is expected to result in an accumulation of  
neutralinos around the core of the Earth, where they will annihilate. An equilibrium density 
is reached when the capture rate equals the annihilation rate. 
Neutrinos are produced in the decays of the resulting particles, with
an energy spectrum extending over a wide range of values and bounded from above by the 
neutralino mass.  Annihilation of neutralinos directly into 
neutrinos (or light fermion pairs in general) is suppressed 
by a factor $m_{\small\mbox{f}}^2/m_{\chi}^2$ due to helicity 
constraints, where $m_{\small\mbox{f}}$ is the fermion mass. \par
 Neutrino detectors can therefore be used to constrain the parameter space 
of supersymmetry by setting limits on the flux of neutrinos from 
 the center of the Earth~\cite{Jungman:96a,Feng:00a}. 
Note that this indirect neutralino detection   
will be favored for high neutralino masses, since the cross section 
of the resulting neutrinos with ordinary matter scales with E$_{\nu}$.

\section{\label{sec:detector}The AMANDA-B10 detector}

 The AMANDA-B10 detector consists of an array of 302 optical modules deployed in  
ten vertical strings at depths between 1500~m and 2000~m in the South Pole 
ice cap. The strings are arranged in two concentric 
circles of 60~m and 120~m diameter respectively. The modules on the 
four inner strings are separated by 20~m in the vertical direction, while 
in the outer six strings the vertical separation between modules is 10~m. 
An optical module consists of a photomultiplier tube housed in a
spherical glass pressure vessel. 
Coaxial cables (in the inner 4 strings) and twisted quad cables (in the outer 6 strings)
provide the high voltage to the photomultiplier tubes and transmit the signals to
the data acquisition electronics at the surface.\par

 Muons from charged-current high-energy neutrino interactions near the array are 
detected by the Cherenkov light they produce when traversing the
ice. The relative timing of the 
Cherenkov photons reaching the optical modules allows the
reconstruction of the muon 
track. A more detailed description of the detector is given in
Ref.~\onlinecite{Amanda:00a}. 
The detector was triggered when a majority requirement was satisfied: an event was recorded if at least  
16 modules had a signal within a predefined time window of 2$\mu$s. 
The data taking rate was 100~Hz.\par

AMANDA-B10 was in operation during the 1997 Antarctic winter. The
separation of 300 atmospheric neutrinos from
the data sample collected in that period established the detector 
as a high-energy neutrino telescope~\cite{Amanda:01a}. The array was upgraded with 
122 more modules during the antarctic summer 1997/1998 and in
1999/2000 253 additional ones were added, completing 
the proposed design of 677 optical modules in 19 strings, AMANDA-II~\cite{Amanda:01b}.

\section{\label{sec:simulations} Signal and Background simulations}
\subsection{Simulation of neutralino annihilations}

Neutralinos can annihilate pair-wise to, e.g., $\ell^{+} \ell^{-}$,
$\mbox{q}\bar{\mbox{q}}$, W$^{+}$W$^{-}$, Z$^0$Z$^0$, 
H$^0_{1,2}$H$^0_{3}$,
Z$^0$H$^0_{1,2}$ and W$^{\pm}$H$^{\mp}$. 
Neutrinos are produced in the decays of these annihilation
products. Neutrinos produced in quark jets (from e.g.\ $\mbox{b}\bar{\mbox{b}}$ or
Higgs bosons) typically have lower energy than those produced from
decays of $\tau$ leptons and gauge bosons. We will refer to the first 
type of annihilation channels as ``soft'' and to the second as
``hard''.\par

The simulations of the expected neutralino signal were done in the 
framework of the SUSY models described in Ref~\onlinecite{Bergstrom:98a}. 
The hadronization and decay of the annihilation products have been 
simulated using {\texttt{PYTHIA}}~\cite{PYTHIA}. The simulations were
performed for six  
different WIMP masses between 10~GeV and 5000~GeV. For each mass, six different
annihilation channels (c$\bar{\mbox{c}}$, $\mbox{b}\bar{\mbox{b}}$, $\mbox{t}\bar{\mbox{t}}$,
$\tau^{+}\tau^{-}$, W$^{+}$W$^{-}$ and Z$^0$Z$^0$) were considered, 
with $1.25\times 10^{6}$ events generated for each. Note that the decay  
of $\mbox{b}$-- and $\mbox{c}$--hadrons will take place in matter instead of
vacuum. This was incorporated in the simulations in an effective 
manner justified by the fact that, for the neutralino masses 
considered, the re-interactions of these heavy  hadrons  with the 
surrounding medium are not dominant, and can be parametrized as an effective 
energy loss at the time of decay. 
As a reference soft spectrum, we chose the annihilation into b$\bar{\mbox{b}}$, 
and as a reference hard spectrum, the annihilation into W$^{+}$W$^{-}$. For 
a given mass, these two spectra can be regarded as extreme cases. We have used 
these channels in the analysis described below, bearing in mind that a typical
 spectrum would lie somewhere in between.\par

\subsection{Simulation of the atmospheric neutrino flux}
 
 Neutrinos from the decay of secondaries produced in  
cosmic ray interactions in the atmosphere constitute the physical  
background to the neutralino search.  We have simulated this
atmospheric neutrino flux using the
  calculations of Lipari~\cite{Lipari:93a}. To obtain the rate of neutrino 
interactions producing muons we have used the neutrino and 
anti-neutrino$-$nucleon cross sections from Gandhi {\em et al.}~\cite{Gandhi:96a}. 
The actual neutrino-nucleon interactions have been simulated with
{\texttt{PYTHIA}} using the {\texttt{CTEQ3}}~\cite{CTEQ3} 
parametrization of the nucleon structure functions. 
 The use of {\texttt{PYTHIA}} allows to model the hadronic
shower produced at the vertex of the interaction and, therefore, to
calculate the Cherenkov light produced by secondaries. When the 
neutrino-nucleon interaction occurs within the instrumented volume of
the detector, this is a non-negligible contribution to the total event 
light output.\par

 A three-year equivalent atmospheric neutrino sample with energies between
10~GeV and 10~TeV and zenith angles between 90$^\circ$ (horizontal)
and 180$^\circ$ (vertically up-going) has been
simulated~\cite{Dalberg:99a}. The sample contains 3.7$\times 10^{7}$
events, of which 41234 triggered the detector. 

\subsection{Simulation of the atmospheric muon flux}

 The majority of the triggers in AMANDA are induced by muons 
produced in cosmic ray interactions in the atmosphere and 
reaching the detector depth. The simulation of this atmospheric 
muon flux was performed using the {\texttt{BASIEV}} \cite{BASIEV} program. 
We note that this program only uses protons as primaries. 
However, the systematic uncertainty introduced by this approximation
is negligible in comparison with that from the present uncertainty in the primary flux 
intensity. Moreover, heavier nuclear primaries  produce more muons per interaction, but 
with lower energies on average~\cite{Gaisser:90a}, which will in general loose all their 
energy and decay before reaching the detector. A study performed using the 
{\texttt{CORSIKA}}~\cite{CORSIKA} air shower generator, with the QGSJET 
option to model the hadronic interactions,  
including the complete cosmic ray composition confirms this 
scenario.\par

 The simulation of a statistically significant sample of atmospheric muon
background is an extremely high CPU-time consuming task due to the
strong rejection factors needed. We have simulated 6.3~$\times$~$10^{10}$  
primary interactions, distributed isotropically with zenith angles, 
$\Theta$, between 0 and 85 degrees,
and with energies, E,  between 1.3~TeV and 1000~TeV, 
assuming a differential energy distribution $\propto
E^{-2.7}$(Ref.~\onlinecite{PDG:00}). The total number of triggers 
produced were 5~$\times$~$10^{6}$. Normalizing to the primary cosmic ray rate, 
the generated  sample corresponds to about 0.6 days of equivalent detector live-time. 
Due to the narrow vertical angular cones used for this analysis this background 
sample is sufficient to model the detector response and develop the 
rejection cuts. In addition, a larger sample  
of background data was used in the training of the discriminant
analysis program used as cut level 4. This is described in more detail in the 
next section. 

\begin{figure}[ht]
\centering\epsfig{file=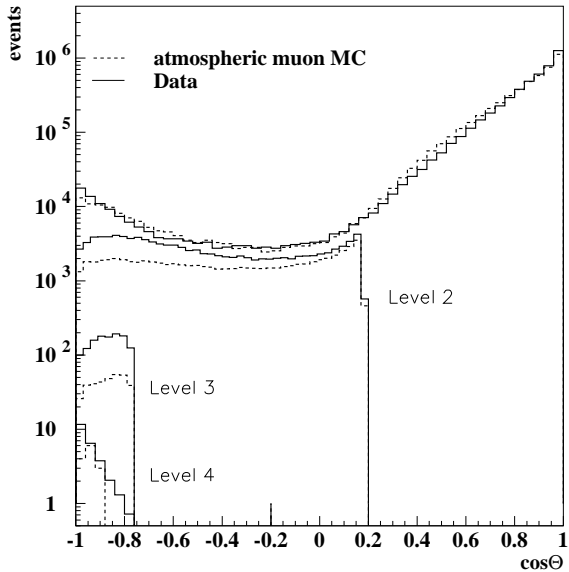,height=\linewidth,width=\linewidth}
\caption{Angular distributions of data and
atmospheric muon simulation Monte Carlo (MC) at different analysis
levels. Top to bottom: trigger to level 4. The
distributions are normalized to the simulated sample, 5$\times$10$^6$ events.}
\label{fig:atm_mu-data_costh_L0L4}
\end{figure}

\subsection{Muon propagation}

The muons produced in the signal and background simulations described above  
were propagated from the production point to the  detector 
taking into account energy losses by bremsstrahlung, pair production, photo-nuclear 
interactions and $\delta$-ray production from~Ref.~\onlinecite{Lohmann:85a}. 
The Cherenkov light emitted by 
the secondaries produced in these processes is taken into account when
calculating the response of the detector to the passage of the muon.\par

\begin{figure}[t]
\centering\epsfig{file=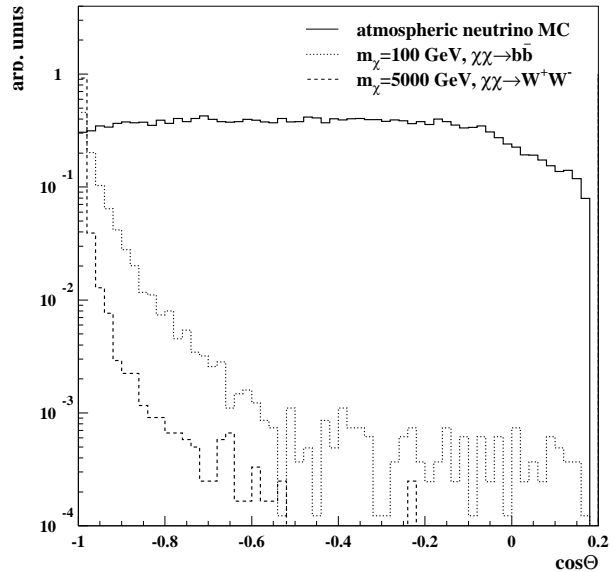,height=\linewidth,width=\linewidth}
\caption{Angular distribution of muons from atmospheric neutrinos and
from the annihilation of neutralinos after filter level 2. The two
extreme neutralino masses and annihilation channels considered in this
paper are shown. The relative normalization is arbitrary.}
\label{fig:wimp-atm_nu_costh_L2}
\end{figure}

\section{\label{sec:analysis}Data analysis}

\begin{figure}[t]
\centering\epsfig{file=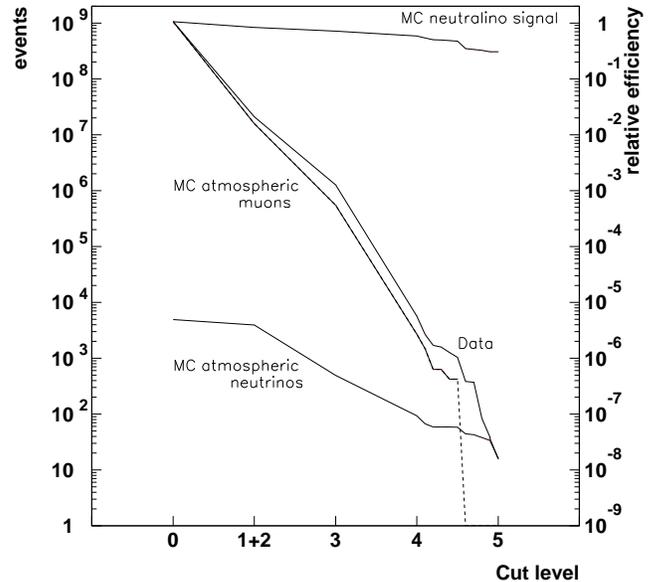,height=\linewidth,width=\linewidth}
\caption{Rejection and efficiency at each filter level for the data
and simulations of the neutralino signal, atmospheric neutrinos and atmospheric muons. 
The dashed part corresponds to rejection levels surpassing the statistical
precision of the simulated sample, yielding zero remaining events.
The neutralino signal curve should be read only with respect to the
right axis scale, and it shows the relative signal efficiency with
respect to trigger level. The rest of the curves are plotted with respect to the
left axis scale.}
\label{fig:cut_efficiencies}
\end{figure}

 The analysis presented in this paper was performed on data taken with 
the 10-string AMANDA detector between March and November 1997. The
experimental data set consists of 1.05$\times$10$^9$ events in a total of 130.1 days of detector 
live-time. The data were first cleaned of noise hits and hits from 
optical modules that were unstable during the running period. 
Short pulses, that are likely induced by cross talk between channels 
are also rejected at this stage. Details on the data cleaning procedure are given 
in~Ref.~\onlinecite{Amanda:00b}. The data is then reconstructed and five filters consisting of  
cuts based on the event hit pattern and the quality of the reconstruction are applied in 
order to identify potential up-going neutrino candidates. 
The distributions of the reconstructed zenith angle from trigger level
(after hit cleaning) until filter level 4 for data 
and simulated atmospheric muons are shown in figure~\ref{fig:atm_mu-data_costh_L0L4}. 
The curves have been normalized to the simulated sample, 5$\times$10$^6$ events. 
The uppermost curves in the plot show the reconstructed direction without any quality criteria 
applied to the fits, showing good agreement between the data and the Monte Carlo sample 
along the whole angular range. The curves clearly indicate that a small percentage (about 2\%) of the
originally down-going tracks are misreconstructed as up-going (cos$\,\Theta$ less than zero the figure). 
The series of cuts described below were developed to reject such
misreconstructions, and their effect on the angular distribution is
also shown in figure~\ref{fig:atm_mu-data_costh_L0L4} for comparison. 
The filter level 2 and level 3 curves show that the filtering
procedure is more effective rejecting the simulated muon background than the data. 
This is due to detector effects not included in the simulation of the detector response and surviving 
to these levels, like electronic cross talk between channels or inefficiencies of the digitizing 
electronics. Other processes not included in the background simulations that can contribute to the 
discrepancy are overlapping events from uncorrelated cosmic ray interactions and the contribution from 
electron neutrino induced cascades. 
 To account for this different behavior between data and simulated background under standard cuts, 
we have used an iterative discriminant analysis as cut level 4 (see subsection~\ref{sec:L4}) 
trained on a sub-sample of data (which represents the real remaining background 
better than the simulations) and a sub-sample of the neutralino signal. 
A final series of high quality cuts were applied after the discriminant analysis, bringing the remaining 
data sample to agree with the number of events expected from the known atmospheric neutrino flux, as 
shown in figure~\ref{fig:cut_efficiencies} and table~\ref{tab:cuts}. 
Note that the atmospheric neutrino curve and the data curve in figure~\ref{fig:cut_efficiencies} join 
and follow each other in the last two steps of the cuts applied within the level 5 filter. 
The next subsections give a more 
detailed description of the variables used and the cuts applied at each filter level.

\subsection{\label{sec:L1}Filter level 1}

 In a first stage, a simple and computationally fast filter  
based on fitting a line to the time pattern of the events, 
was applied to the data sample in order to reject obvious down-going
tracks. This ``line fit'' (LF) assumes that the known  
space point of each hit optical module, $\vec{r}_i$,  is related to the measured hit 
time, $t_i$, by $\vec{r}_i\,=\,\vec{r}_o+\vec{v}t_i$. The minimization of 
$\chi^2\,=\,\sum_{i}(\vec{r}_i-\vec{r}_o-\vec{v}t_i)^2$, where the
index runs over all the hits in the event, leads to an explicit solution for 
$\vec{v}$. The zenith angle of the fitted track is readily obtained as 
$\cos\Theta_{\mbox{\tiny LF}}=-v_z/|v|$. 
 The angular resolution of the line fit is relatively low since it 
does not incorporate any information about the 
geometry of the Cherenkov cone or about scattering of the Cherenkov
photons in the ice. Still, its simplicity and 
computational speed makes it a very useful tool for a first 
assessment of the track direction and for rejection of down-going
atmospheric muons~\cite{Stenger:90a}.  The first level filter rejected 
obvious down-going atmospheric muons by requiring 
 $\Theta_{\mbox{\tiny LF}} > $ 50$^\circ$. 

\begin{figure}[t]
\centering\epsfig{file=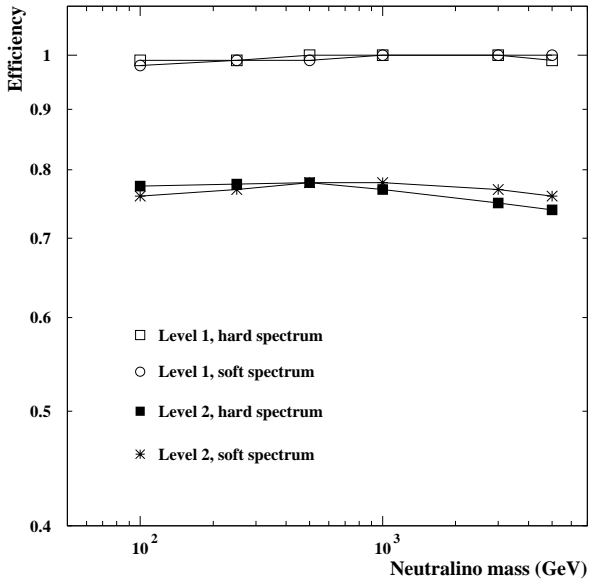,height=\linewidth,width=\linewidth}
\caption{Efficiencies relative to trigger level at filter levels 1 and 2 as a function of the 
neutralino mass.}
\label{fig:L1L2_efficiencies}
\end{figure}

\subsection{\label{sec:L2}Filter level 2}

 The events that pass the level 1 filter are reconstructed using a
maximum likelihood approach (ML) as described in~\cite{Amanda:00a}. 
In short, the ML technique uses an iterative process to maximize the
product of the probabilities that the optical modules receive a signal  
at the measured times, with the track direction
(zenith and azimuth angles) as free parameters. The expected 
time probability distributions include the scattering and 
absorption characteristics of the ice as well as the distance and 
relative orientation of the optical module with respect to the track~\cite{Wiebusch:99a}.\par

 The level 2 filter consists of two cuts: the ML-reconstructed zenith
angle must be larger than 80$^\circ$ and at least three hits must be ``direct''. 
A hit is defined as direct if the time residual, $t_{\mbox{\tiny res}}$ 
(the difference between the measured time and the expected time 
assuming the photon was emitted from the 
reconstructed track and did not suffer any scattering), is
small. Unscattered photons preserve the timing information. Therefore, 
the reconstruction of the direction of tracks with several direct hits 
presents a significantly better angular resolution. 
The number of direct hits associated with a track is the first 
quality requirement applied to the reconstructed data and simulated
samples~\cite{Amanda:00b}. A residual time interval between 
-10~ns and 25~ns was used to classify a hit as direct at this level.\par 

Figure~\ref{fig:wimp-atm_nu_costh_L2} 
shows the zenith angle distributions of simulated muon tracks from neutrinos produced in 
annihilation of neutralinos for the two extreme masses used in this
analysis as compared to that from atmospheric neutrinos after filter level 2. 
The corresponding curve for data and simulated atmospheric muons is included 
in figure~\ref{fig:atm_mu-data_costh_L0L4}. The combined effect of these two filters on the data is a 
rejection of 98\%, as shown in table~\ref{tab:cuts}.  
The efficiencies with respect to trigger level of both level 1 and level 2 filters 
for simulated neutralino signal are shown in figure~\ref{fig:L1L2_efficiencies}, for different 
neutralino masses and the two extreme annihilation channels used. \par

Filters 1 and 2 are applied in an initial data
reduction common to the different subsequent analyses of the data. 
The rest of the cuts described below were specifically
designed for the WIMP search with the aim of identifying and rejecting misreconstructions
while maximizing signal detection efficiency and background rejection~\cite{Pia:00a}.\par 

\subsection{\label{sec:L3}Filter level 3}

\begin{table*}[t]
\caption{\label{tab:cuts}Rejection of data, of the simulated atmospheric neutrinos and atmospheric-muon
          background samples and efficiency for 
the simulated neutralino signal from trigger level to filter level 5.}
\begin{ruledtabular}
\begin{tabular}{ccccc}  
 Filter Level&   Data&   Atmospheric neutrinos & Atmospheric muons& $\chi\bar{\chi} \rightarrow WW$ \\
          & 130.1 days  &   130.1 days equivalent & 0.6 day equivalent & $m_{\chi}$=250 GeV \\
          &(events)   &(events) &(events) &(\% of trigger level)  \\\hline 
   0      & 1.05$\times 10^9$  & 4899 & 5$\times 10^6$ & 100\\ 
   1+2    & 2.3 $\times$ 10$^7$  &  2606    & 7$\times$10$^4$  & \;79  \\  
   3      &  1.2 $\times$10$^6$  & \;472 & 2588        &  \;68   \\ 
   4      &     5441      & \;\;89     & \;\;\;13      &  \;56    \\ 
   5      &  14    & \;\;\;16.0     &  \;\;\;\;0    &  \;29   \\ 
\end{tabular}
\end{ruledtabular}
\end{table*}

The angular distribution of the 
events is the most obvious difference between the predicted neutralino 
signal and both the atmospheric neutrino flux and the atmospheric 
muon background. Neutrinos from neutralino annihilations in the center of the Earth 
 would be concentrated in a narrow cone close to the vertical
direction, while atmospheric neutrinos are distributed
isotropically. The level 3 filter further restricted the
ML-reconstructed zenith angle to be larger than 140$^\circ$, placed a cut  on the total
number of hit modules in the event, N$_{\mbox{\tiny ch}}>$ 10, and on the summed hit probability 
of the modules with a signal, P$_{\mbox{\tiny hit}}>$0.23. The number of  hits with time residuals 
between -10~ns and 25~ns was required to be larger than 4 and 
the number of hits  with residuals between -15~ns and 75~ns to be larger than 5. 
At this stage the possible correlations between the variables are ignored, and the 
cuts applied to each of them individually. Table~\ref{tab:cuts} 
shows the efficiency and rejection power at this cut level. 
Only 5$\times$10$^{-4}$ of the simulated atmospheric muon background survive 
this level, compared with 68\%
of the simulated neutralino signal and 10\% of the atmospheric neutrinos.

\subsection{\label{sec:L4}Filter level 4: Iterative Discriminant Analysis}

 To account for possible correlations between the variables and 
perform a multidimensional cut in parameter space, the next filter level  
was based on an iterative non-linear discriminant analysis, using 
the IDA program~\cite{IDA}. Given a set of $n$ variables, the program 
builds the ``event vector'' $\mbox{\bf x}^k=(x_1, ..., x_n, x_1^2, x_1x_2, ..., x_1x_n,
x_2^2, x_2x_3, ..., x_n^2)$, where $x_i$ is the value of variable $i$ in 
event $k$. A class of events, the signal or background sample, is characterized by their mean vector 
$\left<\mbox{{\bf x}}_s\right>$ or $\left<\mbox{{\bf x}}_b\right>$, and the mean
difference between the samples is given by the vector 
  $\Delta\mu=\langle {\bf x}_s \rangle-\langle {\bf x}_b \rangle$. 
 The spread of the variables is contained in the variance vectors, 
${\bf \mu}^k_s= \mbox{{\bf x}}^k - \langle \mbox{{\bf x}}_s \rangle$ and 
${\bf \mu}^k_b= \mbox{{\bf x}}^k - \langle \mbox{{\bf x}}_b \rangle$, which are used
to define a variance matrix for 
each class, $V^{s,b}=\sum_{k}^{N_{evts}}{\bf \mu^k}_{s,b} ({\bf
\mu^k}_{s,b})^{\mbox{\tiny T}}$, where $N_{evts}$ 
is the number of events in the signal or background samples and ${\mbox{\small T}}$
denotes the transpose. The problem of 
separating signal from background is transformed into the problem of
finding a hyperplane in event vector space which gives minimum local variance for each
class and maximum separation between classes. This is translated into the 
requirement that the ratio  R=$({\bf a}^{\mbox{\tiny T}}\Delta{\bf \mu})^2/{\bf a}^{\mbox{\tiny T}}\,V{\bf a}$ should be
maximal, where here the variance matrix $V$ is the sum of the 
variance matrices for signal and background and {\bf a} is a vector of 
coefficients to be determined by training the program on a signal and a 
background sample. A target signal efficiency 
and background rejection factor are chosen beforehand. 
The coefficients {\bf a} are determined in an iterative process 
carried out until the specified rejection factor is achieved or a predefined number of 
iterations reached. The coefficients found in this way are used to 
select events from the signal region in the multidimensional parameter
space: each event is characterized by the scalar D=${\bf a}^{\mbox{\tiny T}}\,{\bf x}$
and a cut on D serves as the selection criterion. \par

 Eight variables were used in the training of the  discriminant
analysis program and in the subsequent cuts: 
the velocity of the line fit, the number of direct
hits, the number of modules hit, the number of modules hit in the 
string with the largest number of hits, the number of detector 
layers with a hit\footnote{The detector was divided in eight 
horizontal layers of 65~m depth.}, the extension of the event along the three coordinate
axes, the average hit probability and the probability that the event
time pattern is compatible with that expected from a vertical up-going muon. 
This set of variables includes combined information from the fit 
track parameters as well as the general spatial and temporal topology 
of the event.\par

 Since to a first approximation the data consist of atmospheric muon background, 
seven days of data, evenly distributed along the year, were used as the
background training sample. For the signal training sample, muons from
the simulations of 250~GeV neutralinos annihilating into a hard spectrum were used. 
The combination of a relatively low neutralino mass and annihilation into 
the hard channel was chosen as giving a ``typical'' muon spectrum.  
The target signal efficiency was set to 98\% per iteration and the target 
global background rejection to 1000. The stopping criterion was set to 
9 iterations, based on the fact that further loops would reduce the 
number of events in the training sample to a too low number to be
representative of the whole data set. The rejection of background achieved 
was 220 with respect to cut level 3 since the nine loops were exhausted before 
reaching the desired rejection. The overall signal efficiency
attainable after the training process is then (0.98)$^9$=0.83. 
The effect of the discriminant analysis event selection is shown in
table~\ref{tab:cuts}. It indeed achieves the expected signal efficiency, retaining 
82\% of the signal with respect to the previous cut level. The discrepancy of the expected 
number of atmospheric neutrinos and the number of remaining data events at this level indicates 
that the data sample is still contaminated by poorly reconstructed down-going muons. 
A last cut level was therefore developed to improve the rejection of 
the remaining misreconstructed events and select the truly up-going tracks. 

\subsection{\label{sec:L5}Filter level 5: Final event selection}

 The remaining events after the discriminant analysis with a zenith
 angle larger than 165$^\circ$ were 
passed through the following series of cuts.  
The length spanned by the direct hits when projected along the track direction was
required to be at least 110~m, and the vertical length containing all 
hits was required to be at least 170~m.
The $\mbox{z}$ component of
the center of gravity of the direct hits 
(z$_{\mbox{\tiny c.o.g.}}=\sum_{{\mbox{\tiny i}}}{\mbox{z}}_{\mbox{\tiny i}}/N_{\mbox{\tiny direct hits}}$, 
where the sum is over all the direct hits in the event)  
was required to be deeper than 1590~m, and the percentage of hits in the lower half of the
detector less than 55\%. These cuts reject events with a spatially uneven
concentration of hits, typically due to down-going atmospheric muons
that pass just outside the detector or stop close to the
array. \par

\begin{figure*}[t]
\begin{minipage}[t]{0.46\linewidth}
\centering\epsfig{file=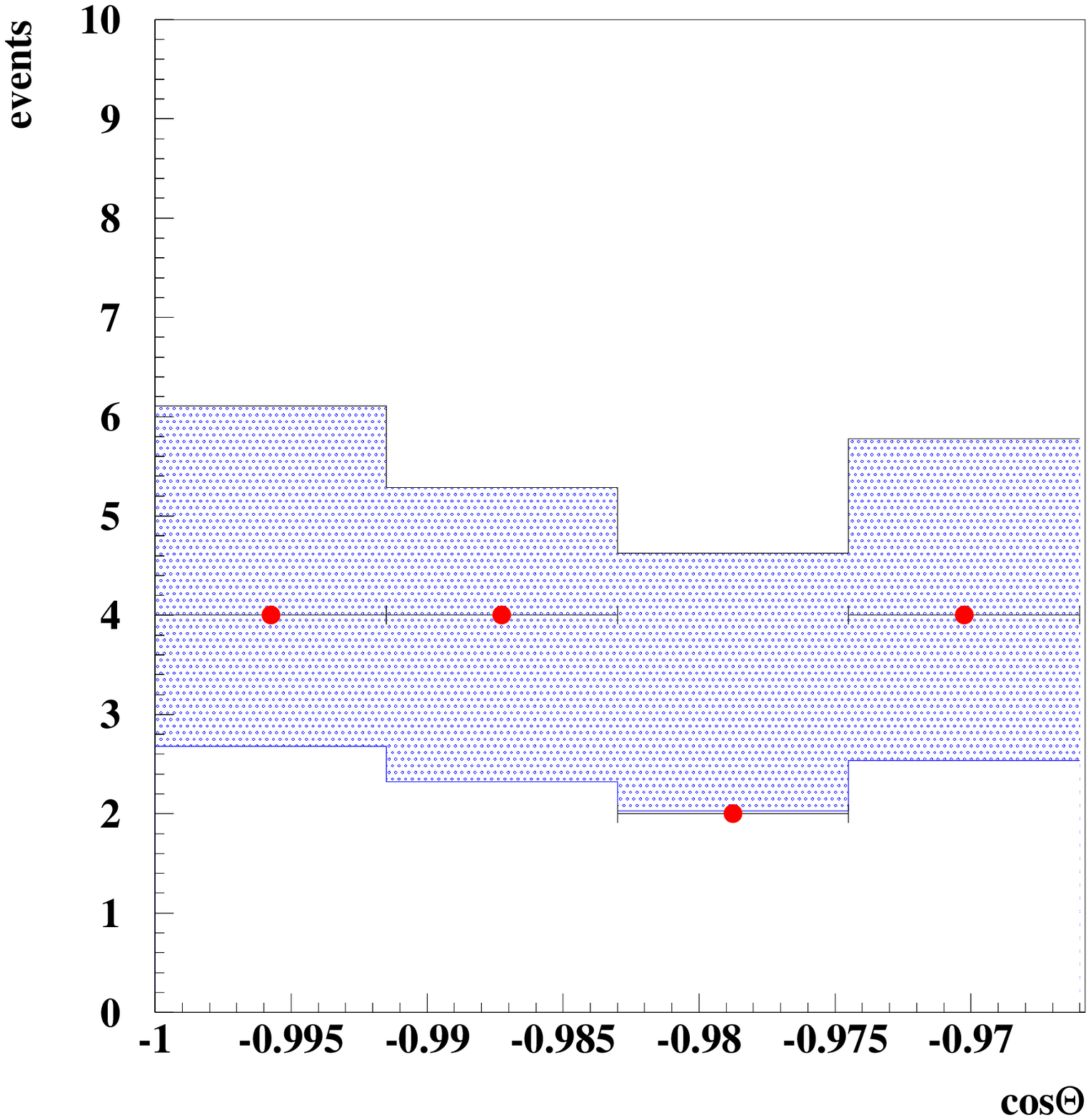,height=\linewidth,width=\linewidth}
\caption{Angular distribution of the remaining data events (dots) 
and simulated atmospheric neutrino events (shaded area) at filter level
5. The angular range shown is between 165$^\circ$ and 180$^\circ$. 
The shaded area represents the total uncertainty in the expected number of events.}
\label{fig:L5_costh_data}
\end{minipage}\hfill
\begin{minipage}[t]{0.46\linewidth}
\centering\epsfig{file=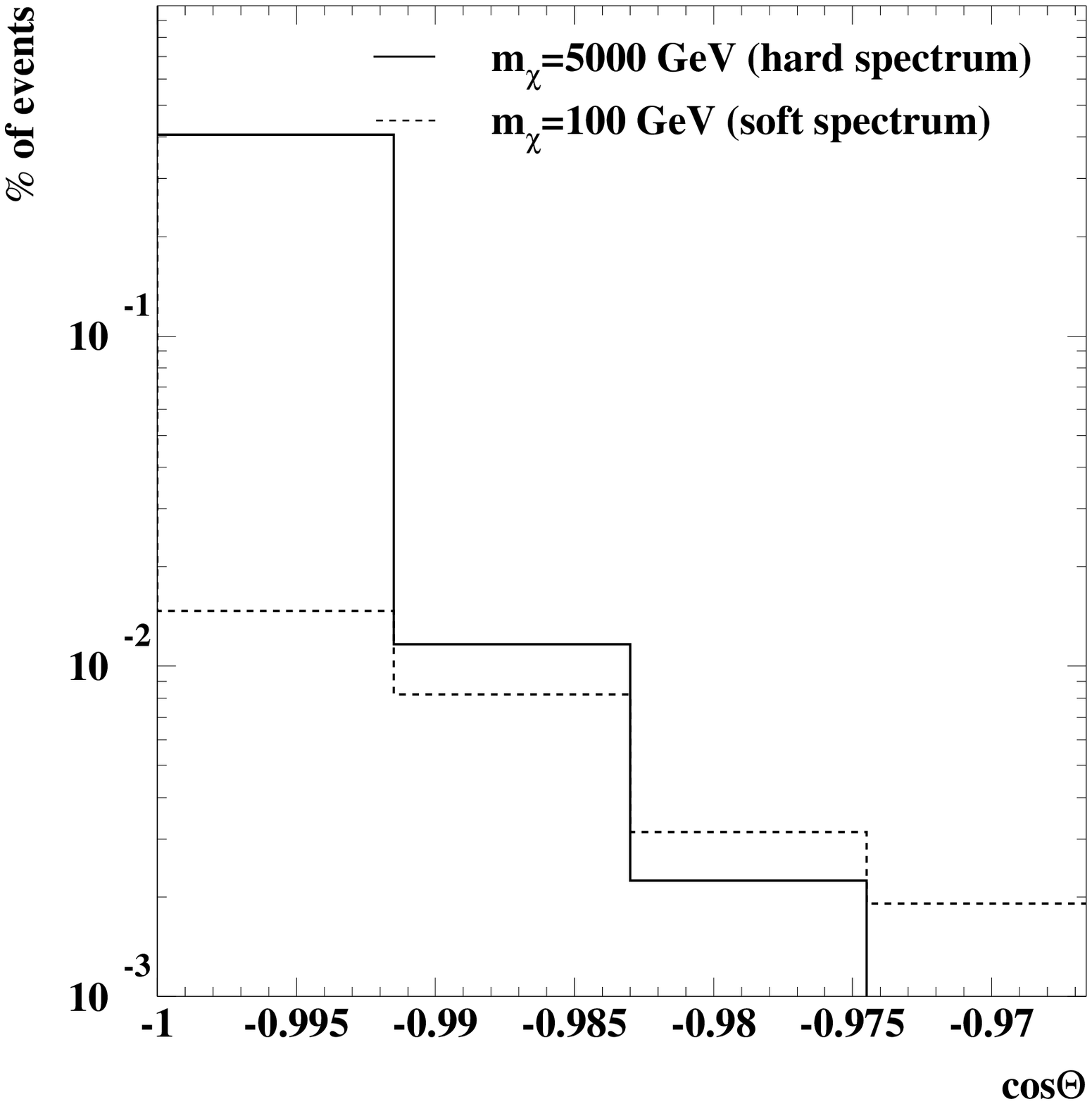,height=\linewidth,width=\linewidth}
\caption{ Angular distribution of the remaining fraction of
neutralinos at filter level5 with respect to trigger level from the two extreme
neutralino masses studied in this paper. The angular range shown is
between 165$^\circ$ and 180$^\circ$.}
\label{fig:L5_costh_wimp}
\end{minipage}
\end{figure*}

 The remaining data at this level  
are consistent with the expected atmospheric neutrino flux. Figure~\ref{fig:L5_costh_data} 
shows the angular distribution of the remaining 14 data events and 
the remaining 16.0 simulated atmospheric neutrino events.  
The angular range shown is for $\Theta>$165$^\circ$, the region where 
a possible neutralino signal is expected to be concentrated. 
No statistically significant discrepancies are 
found between the expected number of events and angular distributions 
of the atmospheric neutrino background and the data. This result is also
consistent with the results on atmospheric neutrinos presented in 
Ref.~\onlinecite{Amanda:01a}.\par
 Due to the different angular shapes of the neutralino signal for
different neutralino masses (see figure~\ref{fig:L5_costh_wimp} for
the two extreme cases considered), we have chosen to restrict further
in angle the signal region we use
to extract the limit on an excess muon flux. We use angular cones that
contain 90\% of the signal for a given neutralino mass. The remaining
data and simulated atmospheric neutrino background events for the 
different angular cones used are shown in table~\ref{tab:evts}.
The background rejection power and signal efficiency from filter level
1 to 5 are shown in figure~\ref{fig:cut_efficiencies} along with
the effect on the data sample.

\section{\label{sec:systematics}Systematic uncertainties}

\begin{figure*}[t]
\begin{minipage}[t]{0.46\linewidth}
\centering\epsfig{file=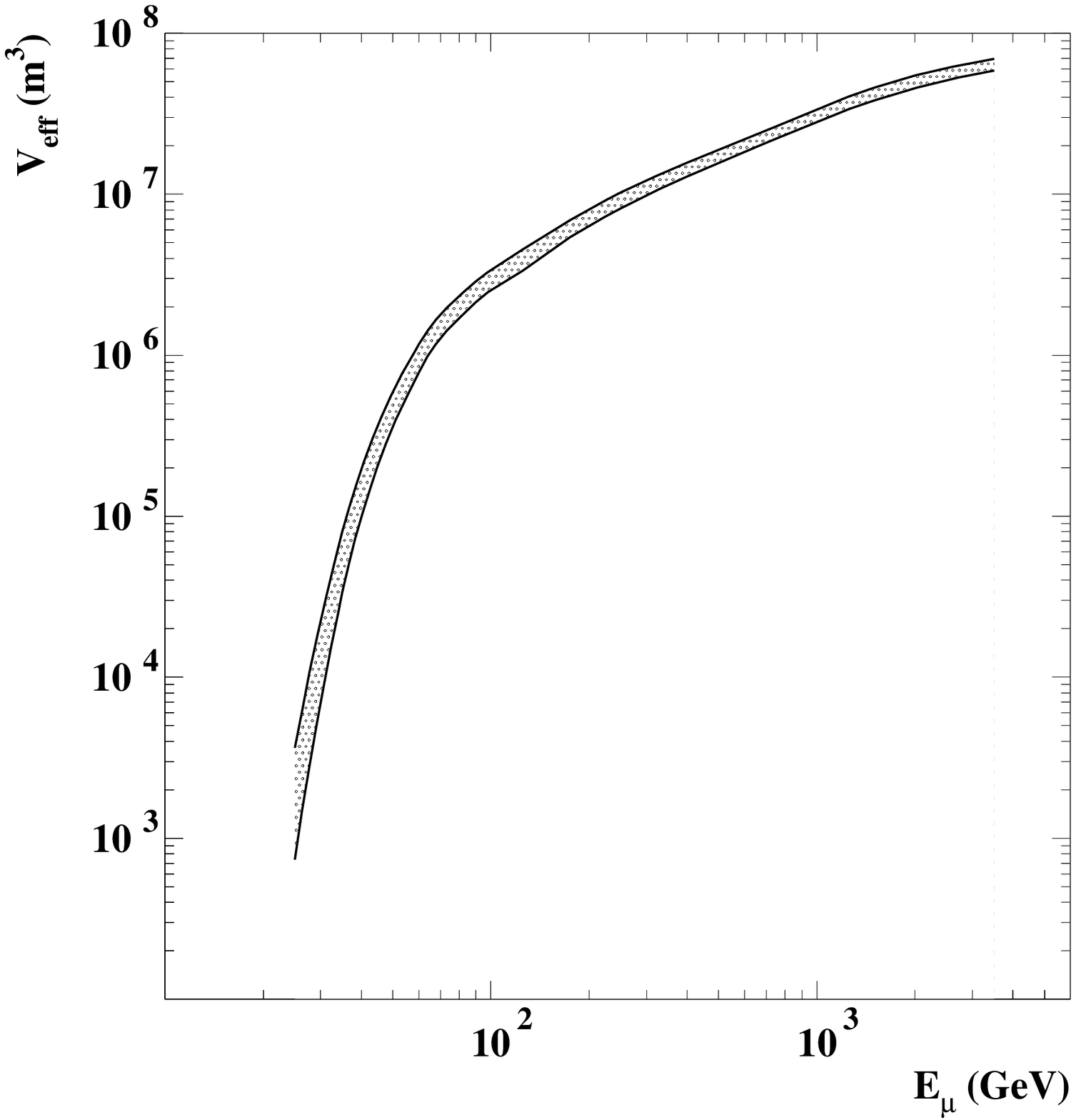,height=\linewidth,width=\linewidth}
\caption{\label{fig:V_eff_vs_E_mu} Effective volume of the detector as
a function of muon energy at filter level 5.}
\end{minipage}\hfill
\begin{minipage}[t]{0.46\linewidth}
\centering\epsfig{file=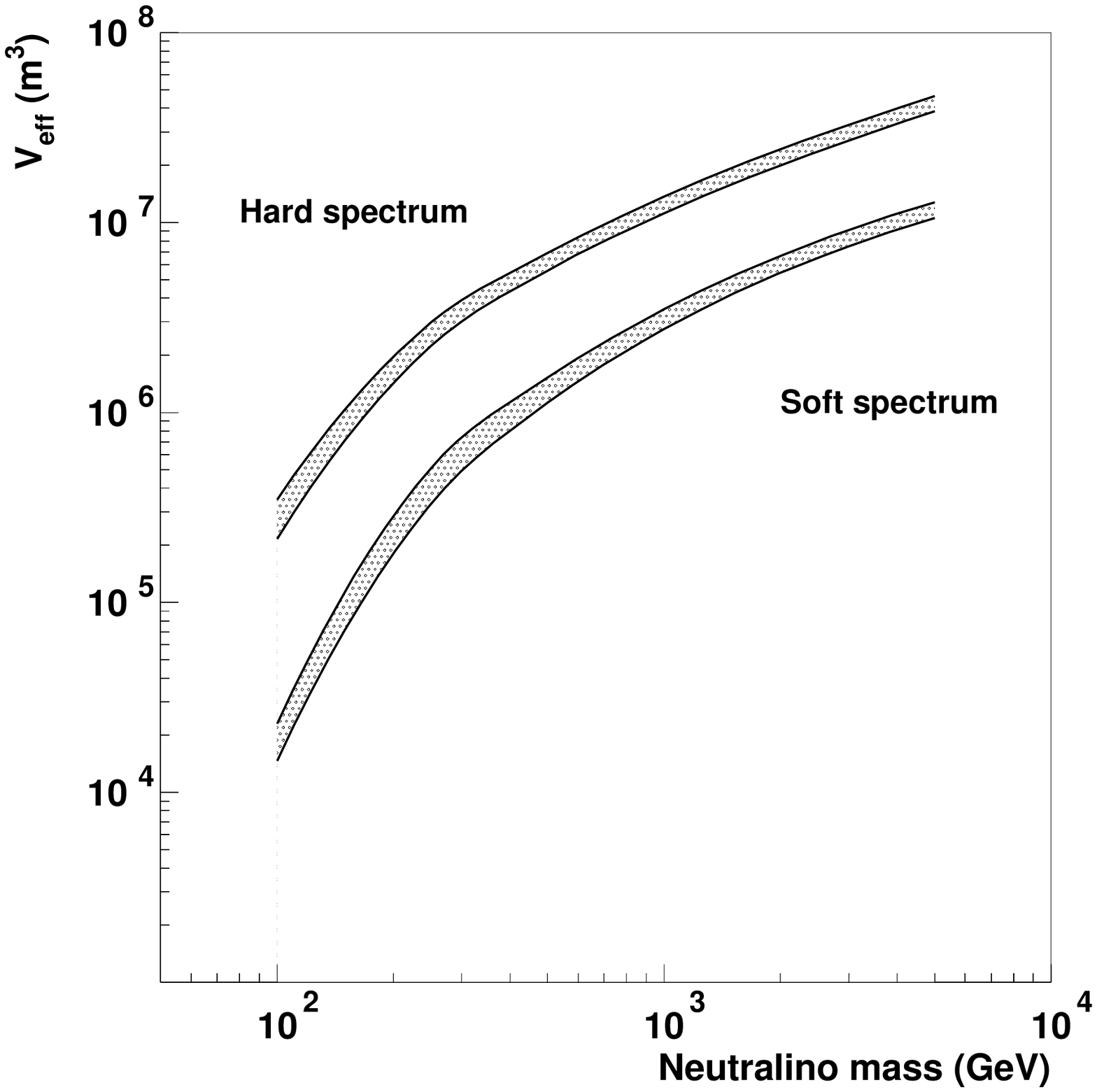,height=\linewidth,width=\linewidth}
\caption{\label{fig:V_eff_L5_mass}Effective volumes for the neutralino
signal as a function of the neutralino mass. }
\end{minipage}
\end{figure*}

 An essential quantity when deriving limits, as we do in the next section,
is the effective volume, $V_{\mbox{\tiny eff}}$, of the detector. It
is the measure of the efficiency to a given signal and it is defined as

\begin{equation}
V_{\mbox{\tiny eff}}\,=\,\frac{n_{\mbox{\tiny L5}}}{n_{\mbox{\tiny gen}}}\,V_{\mbox{\tiny gen}},
\label{eq:V_eff}
\end{equation}

\noindent where $n_{\mbox{\tiny L5}}$ is the number of signal events after
filter level 5 and $n_{\mbox{\tiny gen}}$ the number of events simulated in a volume 
 $V_{\mbox{\tiny gen}}$ surrounding the detector. 
The effective volume of AMANDA-B10 as a function of muon energy 
is shown in figure~\ref{fig:V_eff_vs_E_mu}. 
Given a MSSM model producing a muon flux with a given muon energy spectrum, the effective 
volume of the detector for this particular signal is also calculated through 
equation~\ref{eq:V_eff}. This is shown in
figure~\ref{fig:V_eff_L5_mass} for the different neutralino masses 
used in this analysis. 
The shaded bands in both figures indicate the  
systematic uncertainty estimated as described below.\par
 
 The evaluation of $V_{\mbox{\tiny eff}}$ is subject 
to experimental and theoretical systematic uncertainties present in
the analysis. We have performed a detailed study of the effect of the
uncertainty in several variables on
the resulting effective volume by propagating variations in any of them to the final 
evaluation of $V_{\mbox{\tiny eff}}$. \par

Measurements of the scattering and absorption lengths, $\lambda_s$ and
$\lambda_a$, using pulsed and DC light sources deployed with the detector at different
depths and YAG laser light sent from the surface through optical fibers, have shown that 
these quantities exhibit a depth dependence which is correlated
with dust concentration at different levels in the
ice~\cite{Kurt:99a}. A simulation of the detector response  
including layers of ice with different optical properties has 
been developed and used to evaluate its effect on the results. The
effects introduced are muon-energy dependent and therefore
dependent on the neutralino model. The effective volumes calculated
with the layered ice model are reduced between 1\% and 20\% with respect to the 
uniform ice model, except for the lower neutralino mass and soft
annihilation channel (100 GeV) where the effect reaches 50\%.\par

A further correction accounts for the uncertainties in the optical modules' 
 total and angular sensitivities. It is known that during the 
process of re-freezing after deployment, air bubbles appear in the 
column of ice that has been melted, changing locally the scattering length of
the ice and distorting the effective optical module angular
sensitivity with respect to that measured in
the laboratory. We have used a specific ice model for the ice in the holes
that accommodates this effect. The fact that it appears after 
deployment and that it is not directly measurable in the laboratory makes it 
difficult to assess. Only by an iterative process of comparison of
data and different hole-ice models can it be quantified. We estimate
this effect to yield and increase of 20\% in effective volume with respect to the
uniform angular response model with, again, the soft annihilation
channel of the lowest mass neutralino giving a stronger effect of
34\%. An additional 20\% uncertainty on   
the total optical module sensitivity has been used.\par

 The way to combine of all these effects into a final estimate of the total uncertainty 
in $V_{\mbox{\tiny eff}}$ is a difficult subject, since they are not independent
contributions. As described in the previous paragraphs, by varying the
initial  parameters used in the simulations of the detector and in 
the ice properties, we have obtained a range of possible values for
the effective volume, which we consider as equally probable giving our
current understanding of the detector. 
We have chosen to take the nominal 
$V_{\mbox{\tiny eff}}$ to be used in equation~\ref{eq:V_eff} as the middle value of this range.  
As a conservative estimate of the uncertainty we take half the width
of the range of values obtained. 
We thus conclude that our current estimate of $V_{\mbox{\tiny eff}}$ is affected
by a systematic uncertainty $\sigma_{\tiny V_{\mbox{eff}}}/V_{\mbox{\tiny eff}}$ between 10\% and 25\% 
depending on the neutralino mass considered, the lower mass of 100~GeV giving the larger relative
error. A similar estimate including the same effects has been
made for the atmospheric neutrino Monte Carlo. In this case we estimate the uncertainty on the
effective volume for atmospheric neutrinos to be 20\%.\par

 Further uncertainty in the number of 
expected atmospheric neutrinos (column 3 in table~\ref{tab:cuts}) is
caused by the uncertainties present in the calculation of the atmospheric neutrino flux. 
This is estimated to be of the order of 30\% in the energy region relevant to this 
analysis, and originates mainly from uncertainties in the 
normalization of the primary cosmic ray spectrum and in the hadronic 
cross sections involved~\cite{Gaisser:01a}. This  has been
taken into account as an additional effect on top of the experimental
uncertainty on the effective volume for atmospheric neutrinos, as described in section~\ref{sec:limits2}.\par

It has recently been shown that different muon propagation codes can produce 
differences in the muon flux and energy spectrum at the detector depth (see for
example Ref.~\onlinecite{Sokalski:01a}). 
The code used in this analysis uses the Lohmann~\cite{Lohmann:85a}
parametrizations for muon energy loss, which produce results in agreement within about 10\% of more
recent codes~\cite{Dima:01a} for muon energies up to a few of TeV. We have not
included any  systematics arising from the treatment of muon propagation in the ice in this analysis. 

\section{\label{sec:results}Results}

 From the observed number of events, n$_{\mbox{\tiny obs}}$, and the number of expected atmospheric
neutrino background events, n$_{\mbox{\tiny B}}$, an 
upper limit on the signal, N$_{\beta}$, at a chosen confidence level
$\beta$\%, can be obtained. We have used the unified approach for confidence belt
construction~\cite{Stuart:91a} to calculate 90\% confidence level
limits. In section B below we briefly describe a novel way of calculating 
limits in the presence of systematic uncertainties that we have used 
to obtain the final numbers presented in this paper.\par

\begin{table}[t]
\caption{\label{tab:evts} Number of data events, simulated atmospheric
neutrino background events and the corresponding N$_{\mbox{\tiny 90}}$ for the  angular cones
containing 90\% of the signal for the different neutralino masses. These angular cuts are
applied in addition to the level 5 filter described in
section~\ref{sec:analysis}. ``s'' and ``h'' denote the soft and hard
annihilation channels. 
The numbers in parenthesis in column 5 show N$_{\mbox{\tiny 90}}$
obtained without including systematic uncertainties.}
\begin{ruledtabular}
\begin{tabular}{lcccr}  
   \mbox{m$_\chi$} &  \text{Angular cut}&    \text{Data} &\text{Atmospheric} &  \mbox{N$_{\mbox{\tiny 90}}$}\\
   \text{(GeV)} & \text{(deg)}& \text{(evts.)}  & \text{neutrinos (evts.)}& \\\hline 

   \;100s   &167.5  &    10 & 12.1 & 9.2(4.7)\\ 
   \;100h   &168.5  &   \;\;9 & 10.8 & 6.6(4.7) \\  
   \;250s    &170.0   & \;\;7 &  \;\;8.6 & 5.9(4.1)\\ 
   \begin{tabular}{ll}
   250h   &\multirow{2}{3mm}{$\bigg\}$}\\
   500s    &  \\
   \end{tabular}    & 172.0  &\; 5 & \;\;6.1 & 5.6(3.9)\\
   \;1000s&  173.0   &\;\;4 & \;\;4.6 & 5.3(3.9)\\ 
   \;500h   &173.5  &   \;\;4 & \;\;4.6 & 5.3(3.9)\\ 
   \begin{tabular}{ll}
   1000h &\multirow{2}{3mm}{$\bigg\}$}\\
   3000s  &\\ 
   \end{tabular} &174.0  &\;\;4& \;\;3.9 & 5.6(4.7)\\
   \begin{tabular}{ll}
    3000h & \multirow{3}{3mm}{$\bigg\}$}\\
    5000s &  \\
    5000h &  \\
   \end{tabular} & 174.5& \;\;3 & \;\;3.9& 4.4(3.6)\\

\end{tabular}
\end{ruledtabular}
\end{table}

\subsection{\label{sec:limits1}Flux limits: the standard approach}

\begin{figure*}[t]
\begin{minipage}[t]{0.46\linewidth}
\epsfig{file=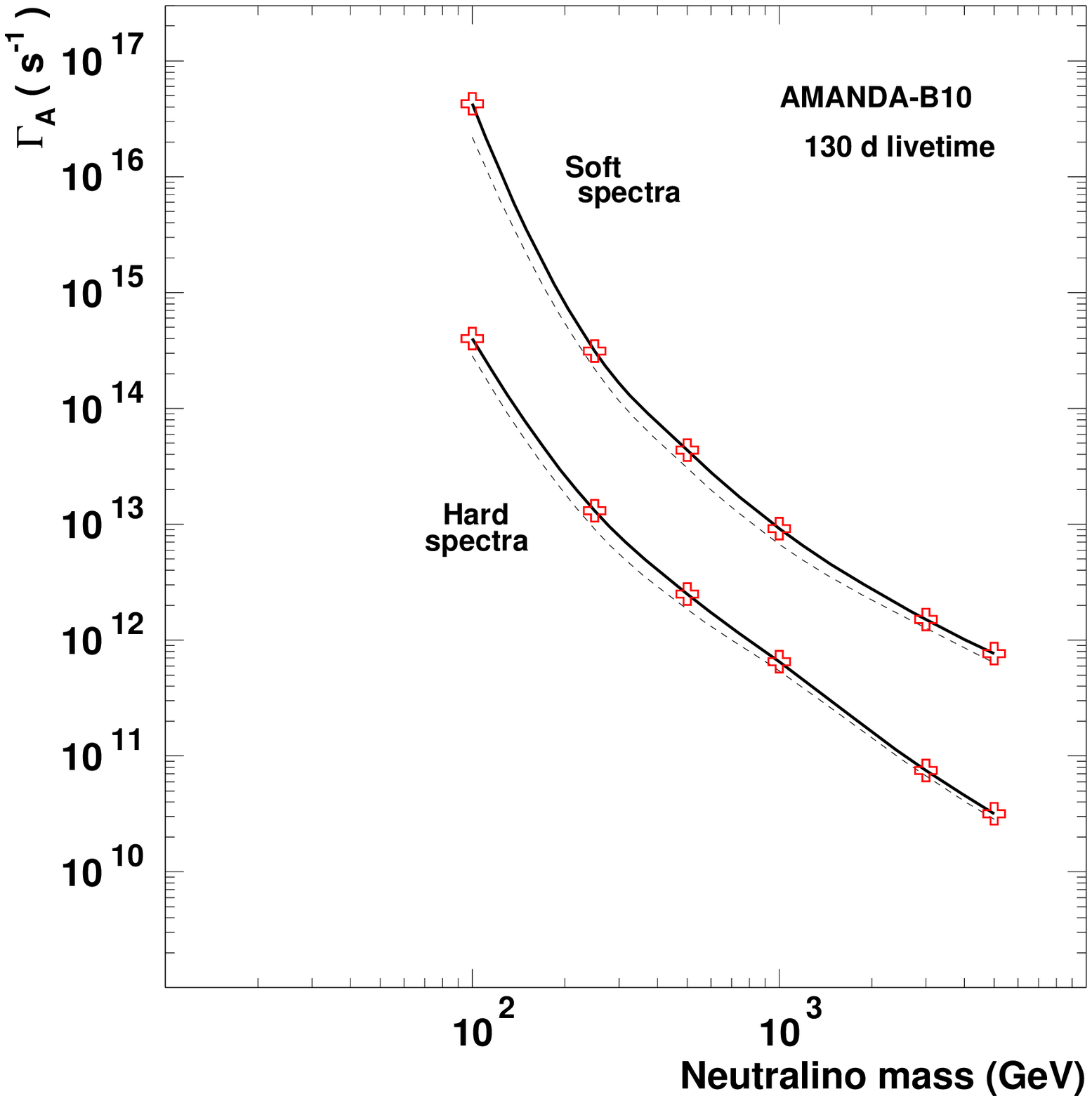,height=\linewidth,width=\linewidth}
\caption{90\% confidence level upper limits on the neutralino 
annihilation rate, $\Gamma_{\mbox{\tiny A}}$, in the center of the Earth as a function of the 
neutralino mass and for the two extreme annihilation channels
considered in the analysis. The dashed lines indicate the limits
obtained without including systematic uncertainties and correspond to the numbers in 
parentheses~in~table~\ref{tab:limits2}. The symbols indicate the
masses used in the analysis. Lines are to guide the eye.}
\label{fig:Gamma_A_limit}
\end{minipage}\hfill
\begin{minipage}[t]{0.46\linewidth}
\epsfig{file=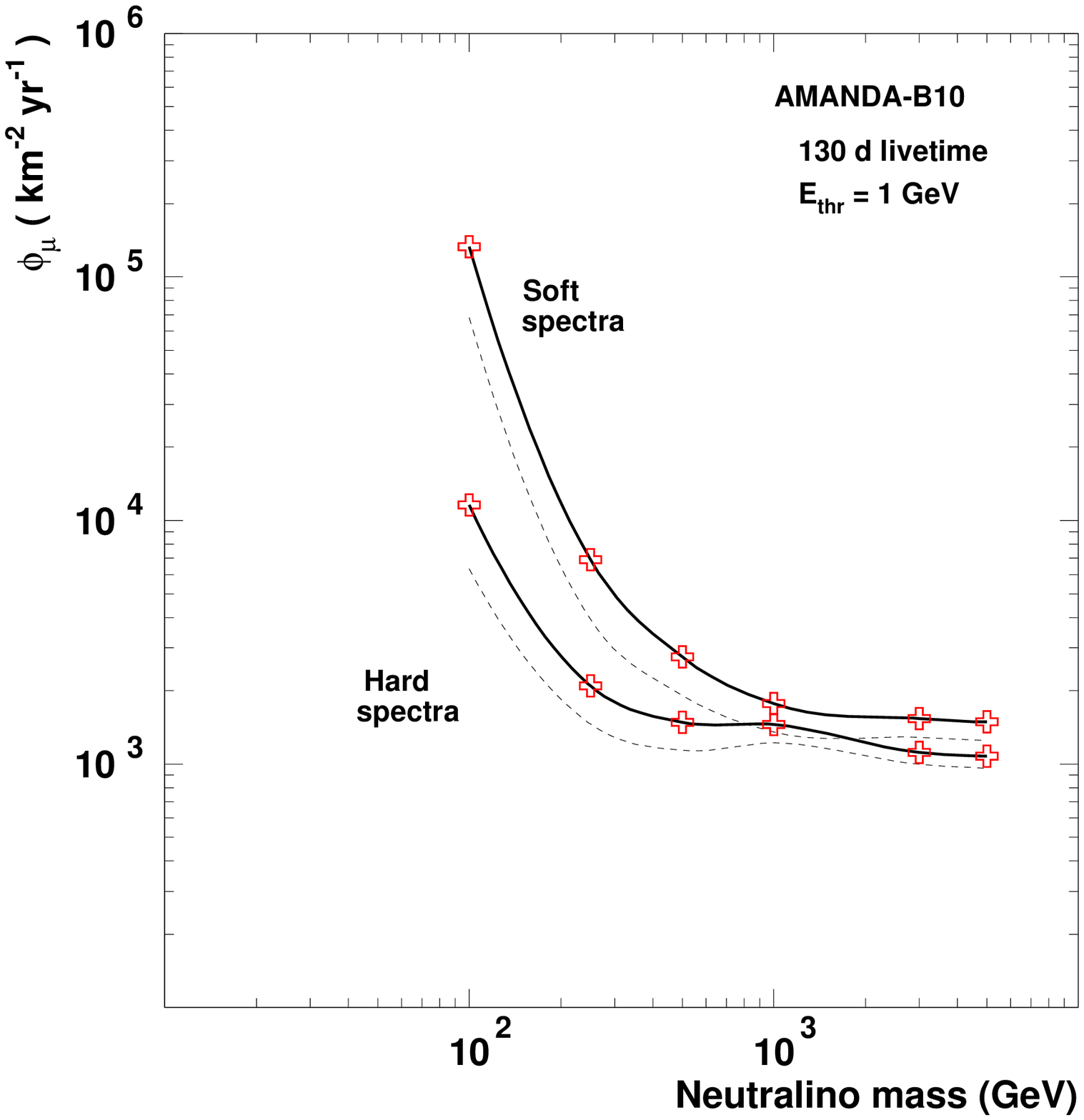,height=\linewidth,width=\linewidth}
\caption{90\% confidence level upper limits  on the muon 
flux at the surface of the Earth, $\phi_{\mu}$, as a function of the 
neutralino mass and for the two extreme annihilation channels
considered in the analysis. The dashed lines indicate the limits
obtained without including systematic uncertainties and correspond to the numbers in 
parentheses~in~table~\ref{tab:limits2}. The symbols indicate the
masses used in the analysis. Lines are to guide the eye.}
\label{fig:flux_limit}
\end{minipage}
\end{figure*}

 For detectors with a fixed geometrical area A, it is natural to derive 
a muon flux limit directly through $\phi_{\mu}\le
\mbox{N}_{\beta}/A\cdot t$, where $t$ is the detector live-time. 
However, due to the large volume of AMANDA and the lack of sharp  
geometrical boundaries it is the effective volume 
$V_{\mbox{\tiny eff}}$, as defined in equation~\ref{eq:V_eff}, that has
to be used to determine a limit on the volumetric neutrino-to-muon conversion rate,
$\Gamma_{\nu\mu}$. The effective
volume provides a measure of the detector efficiency 
since, in addition to through-going tracks, it takes into account the effect of tracks starting or stopping 
within the detector. A limit can then be set on $\Gamma_{\nu\mu}$, that is, on the 
number of muons with an energy above the detector threshold
$E_{\mbox{\tiny thr}}$ produced by neutrino interactions per unit volume and time,

\begin{equation}
 \Gamma_{\nu\mu}\,\le\,\frac{\mbox{N}_{\mbox{\tiny 90}}}{V_{\mbox{\tiny eff}}\cdot t}
\label{eq:Gamma_numu_limit}
\end{equation}
  
$\Gamma_{\nu \mu}$ includes all the detector threshold effects and model dependencies, as indicated
below, and can be directly related to a more physically 
meaningful quantity, the annihilation rate, $\Gamma_{\mbox{\tiny A}}$,
of neutralinos in the center of the Earth through

\begin{table*}[t]
\caption{\label{tab:limits2} The 90$\%$ confidence level upper limits
on the muon flux from neutralino annihilations in the center of the
Earth, $\phi_{\mu}$, for a muon energy threshold $\ge$ 1~GeV. The last column shows the 
threshold-independent neutralino annihilation rate, $\Gamma_{\mbox{\tiny A}}$.
Detector systematic uncertainties have been included in the
calculation of the limits. The corresponding numbers without including uncertainties are
shown in parenthesis}
\begin{ruledtabular}
\begin{tabular}{rrccc}   
$m_{\chi}[GeV]$&annihil.& {\bf $\phi_{\mu} $}
& \multicolumn{1}{c}{\bf $\Gamma_{\mbox{\tiny A}}$}\\
   &channel &$[\times 10^{3} {\rm Km^{-2}yr^{-1}} ]$ & $[{\rm s^{-1}}]$\\\hline\hline
100& hard         & 8.9 \;(6.3)  &  4.0 (2.9)$\times$10$^{14}$  \\
   &soft          &\!\!\!\!\!\!\!\! 133.5\;(68.2) & 4.3 (2.2)$\times$10$^{16}$  \\ \hline
250& hard         & 2.1 \;(1.5)  &  1.3 (0.9)$\times$10$^{13}$  \\
   &soft          & 6.9 \;(3.9)  &  3.8 (2.2)$\times$10$^{14}$  \\ \hline
500& hard         & 1.5 \;(1.1)  &  2.5 (1.8)$\times$10$^{12}$  \\
   & soft         & 2.7 \;(1.9)  &  4.4 (3.0)$\times$10$^{13}$  \\ \hline  
1000& hard        & 1.5 \;(1.2)  &  6.5 (5.4)$\times$10$^{11}$  \\
    & soft        & 1.8 \;(1.4)  &  9.2 (6.8)$\times$10$^{12}$  \\ \hline 
3000&  hard       & 1.1 \;(1.0)  &  7.5 (6.7)$\times$10$^{10}$  \\
    &soft         & 1.5 \;(1.3)  &  1.5 (1.3)$\times$10$^{12}$  \\ \hline
5000& hard        & 1.1 \;(1.0)  &  3.2 (2.8)$\times$10$^{10}$  \\
     &soft        & 1.5 \;(1.2)  &  7.6 (6.4)$\times$10$^{11}$ 
\end{tabular}
\end{ruledtabular}
\end{table*}

\begin{equation}
\begin{split}
\Gamma_{\nu\mu}(m_{\chi} )=\Gamma_{\mbox{\tiny A}} \cdot \frac{1}{4 \pi R_{\oplus}^2} \int_{0}^{m_{\chi}} \sum
B_{\chi{\bar \chi}\rightarrow X}\,\left( \frac{ dN_{\nu}}{dE_{\nu}}
\right) \, & \\
 \times \,\sigma_{_{\nu+N\rightarrow \mu+... }}(E_{\nu}|E_{\mu}\ge E_{\mbox{\tiny thr}}) \,\rho_{\mbox {\tiny N}} \, dE_{\nu},
\end{split}
\label{eq:Gamma_numu}
\end{equation}
\noindent where the term inside the integral takes into account the 
production of muons through the neutrino-nucleon cross 
section, $\sigma_{\nu+N}$, weighted by the different branching ratios 
of the $\chi{\bar \chi}$ annihilation process and the 
corresponding neutrino energy spectra, $B_{\chi{\bar \chi}\rightarrow
X}$ $dN_{\nu}/dE_{\nu}$. $\rho_{\mbox {\tiny N}}$ is the nucleon density of the ice
and $R_{\oplus}$ is the radius of the Earth. We have used a muon
energy threshold of 10 GeV in the simulations of the signal, which has
been taken into account through the muon production cross section.\par

Equation~\ref{eq:Gamma_numu} is solved for $\Gamma_{\mbox{\tiny A}}$. 
$\Gamma_{\mbox{\tiny A}}$ depends on the MSSM model assumptions, as well as the galactic halo model 
used, being related to the capture rate of neutralinos in the Earth. 
Different neutralino models predict different capture and annihilation rates 
that can be probed by experimental limits set on 
$\Gamma_{\mbox{\tiny A}}$. The right column of table~\ref{tab:limits2} shows the limits 
thus derived for $\Gamma_{\mbox{\tiny A}}$. The corresponding curves 
are shown in figure~\ref{fig:Gamma_A_limit}. 
 Quoting limits on the annihilation rate has the advantage that the detector
efficiency and threshold are included through
equation~\ref{eq:Gamma_numu_limit} and, therefore, numbers published by different experiments are directly
comparable. This is not usually the case when presenting limits on  
muon fluxes, where at least the detector energy threshold enters in a 
non trivial way and prevents direct comparison between experiments.  
However, since it is common in the literature to present limits 
on the muon flux per unit area and time, 
we transform below our limit on $\Gamma_{\mbox{\tiny A}}$ into a limit 
on the muon flux from neutralino annihilations in the center of the
Earth. \par

 The total number of muons per unit area and time above any energy threshold
E$_{\mbox{\tiny thr}}$ within a cone of half angle $\theta_c$ as a function of 
the annihilation rate is

\begin{equation}
\phi_\mu(E_\mu \ge E_{\mbox{\tiny thr}},\theta \ge \theta_c)\,=\,
\frac{\Gamma_{\mbox{\tiny A}}}{4\pi R_{\oplus}^2}
\int_{E_{\mbox{\tiny thr}}}^{\infty} dE_\mu
\int_{\theta_c}^{\pi} d\theta
\frac{d^2N_\mu}{dE_\mu d\theta},
\label{eq:mu_flux1}
\end{equation}

\noindent where the term $d^2N_\mu/dE_\mu d\theta$ represents the
number of muons per unit angle and energy produced from the 
neutralino annihilations, and includes all the MSSM model
dependencies for neutrino production from neutralino annihilation and  
the neutrino-nucleon interaction kinematics, 
as well as muon energy losses from the production point to the detector. 
The upper limits on the annihilation rate are thus 
converted to a limit on the neutralino-induced muon flux at any depth and above 
any chosen energy threshold and angular aperture. The 90\% confidence 
level upper limits on the annihilation rate and 
the muon flux at an energy threshold of 1~GeV derived using 
equations~\ref{eq:Gamma_numu_limit},~\ref{eq:Gamma_numu} and~\ref{eq:mu_flux1} 
are shown in parenthesis in table~\ref{tab:limits2}. 
The fluxes have been corrected for the inefficiency introduced by
using angular cones that include 90\% of the signal, so the numbers
presented represent the limit on the total muon flux for each neutralino
model. The threshold of 1 GeV has been chosen to be able to compare with published limits by 
other experiments that have similar muon thresholds (see section~\ref{sec:comparison} below).

\subsection{\label{sec:limits2}Evaluation of the limits including systematic uncertainties}

 However, the best limits an experiment can set are affected by the
systematic uncertainties entering the analysis. Including the known
theoretical and experimental systematic
uncertainties in the calculation of a flux limit is not 
straightforward, and often overlooked in the literature.  
A precise evaluation of a limit should involve the incorporation of 
both the uncertainties in the background counts, $\sigma_{\mbox{\tiny b}}$, 
and in the effective volume, $\sigma_{\mbox{\tiny V}}$. An additional caveat arises since  
the uncertainty in the effective volume introduces in turn an 
additional uncertainty in the expected number of background events, on
top of the 30\% uncertainty used in the background neutrino flux $\sigma_{\mbox{\tiny b}}$. 
A proper implementation of the systematics in the calculation of a
limit should take this correlation into account.

 One approach to incorporate systematic uncertainties into an upper limit 
has been proposed in Ref.~\onlinecite{Cousins:92a}. We have developed a 
similar method suited to our specific case which 
includes the systematic uncertainty in $V_{\mbox{\tiny eff}}$ in the calculation 
of N$_{\mbox{\tiny 90}}$ used in equation~\ref{eq:Gamma_numu_limit}.
 The method is a modified Neyman-type confidence belt construction~\cite{Jan:01a}. 
 The confidence belt for a desired confidence level $\beta$ is
 constructed in the usual way by integrating the Poisson distribution with mean 
  n$_{\mbox{\tiny tot}}$=n$_{\mbox{\tiny S}}$+n$_{\mbox{\tiny B}}$
so as to include a $\beta$\% probability content. But the 
number of events for signal and background, n$_{\mbox{\tiny S}}$ and 
n$_{\mbox{\tiny B}}$, are taken themselves to be random
 variables obtained from Gaussian distributions with  means equal to the
actual number of signal and background events observed and widths 
 corresponding to the systematic uncertainties in signal and
background. \par

 Given an experimentally observed number of events, N$_{\mbox{\tiny exp}}$, 
the 90\% confidence level limit on the number of signal events is obtained by 
simply inverting the calculated N$_{\mbox{\tiny 90}}(\mbox{n}_{\mbox{\tiny tot}})$ at the 
corresponding n$_{\mbox{\tiny tot}}$=N$_{\mbox{\tiny exp}}$ value. 
In this way the  different uncertainties for signal and background and
 the correlation  between them are included naturally. \par

 In summary, the inclusion of our present systematic uncertainties in 
the flux limit calculation yields results which are weakened between $\sim$10\% and 
$\sim$40\% (practically a factor of 2 for the soft channel of
m$_\chi$=100 GeV) with respect to those obtained using N$_{90}$ 
calculated without systematics. The effect is dependent on the
WIMP mass, and it reflects the better sensitivity of AMANDA for higher neutrino energies. 
Figures~\ref{fig:Gamma_A_limit} and \ref{fig:flux_limit} show 
the 90\% confidence level limit on the neutralino annihilation rate
and the corresponding limit on the resulting muon flux for a muon threshold of 1 GeV for 
the hard and soft annihilation channels considered in the analysis. The symbols show the 
particular neutralino masses used in the simulation. The lines are to guide the eye and they show the 
limits obtained including systematic uncertainties (solid). The dashed lines,
included for comparison, show the values obtained using the 
Neyman construction with the unified ordering scheme without including
uncertainties. Table~\ref{tab:limits2} summarizes the corresponding numbers.

\subsection{Effect of neutrino oscillations}
\label{sec:oscillations}
 
 To account for neutrino oscillations among  the different flavors, the 
atmospheric neutrino spectrum should be weighted by a factor W(E$_{\nu}$)
which includes the probability that a muon neutrino has oscillated 
into another flavor in its way through the Earth to the detector. For 
the purpose of illustration consider a two-flavor oscillation scenario between 
$\nu_\mu$ and $\nu_\tau$. Then 
W(E$_{\nu}$) =
$1-\mbox{sin}^2(2\theta)\,\mbox{sin}^2\left(1.27\,\Delta\mbox{m}^2[\mbox{eV}^2]\,
\mbox{D}_{\oplus}[\mbox{km}]/\mbox{E}_{\nu}[\mbox{GeV}]\right)$,
where $\mbox{D}_{\oplus}$ is the diameter of the Earth, $\theta$ the
mixing angle and $\Delta\mbox{m}^2$ the difference of the squares of
the flavor masses. 
 Note that the effect depends strongly on the neutrino energy and it 
is negligible in the high energy tail of the atmospheric spectrum since the 
oscillation length is then much larger than the Earth diameter. 
If we choose $\mbox{sin}^2(2\theta)$=1 and $\Delta\mbox{m}^2=2.5\times 10^{-3}$
eV$^2$ based on the results obtained in Ref.~\onlinecite{Kameda:01a},
the number of expected atmospheric neutrino events is reduced between 5\% and 10\%,
depending on the angular cone considered. This would weaken the limits
by about the same amount.\par

The effect of neutrino oscillations on the possible WIMP signal is 
model dependent and has been estimated in~Refs.~\onlinecite{Marek:01a}
and~\onlinecite{Fornengo:00a}. However the authors reach different 
conclusions on the direction of the effect: up to a factor of two in 
increased muon flux in Ref.~\onlinecite{Marek:01a} and a reduction of 
about 25\% in Ref.~\onlinecite{Fornengo:00a} for a neutralino mass of
100~GeV. For higher neutralino masses both authors predict a less 
pronounced effect, which becomes negligible for the higher masses 
considered in~\cite{Marek:01a} (m$_\chi>$ 300~GeV). We have not
included any oscillation effect on the neutrinos from the WIMP 
signals considered in this paper.
 
\begin{figure}[t]
\centering\epsfig{file=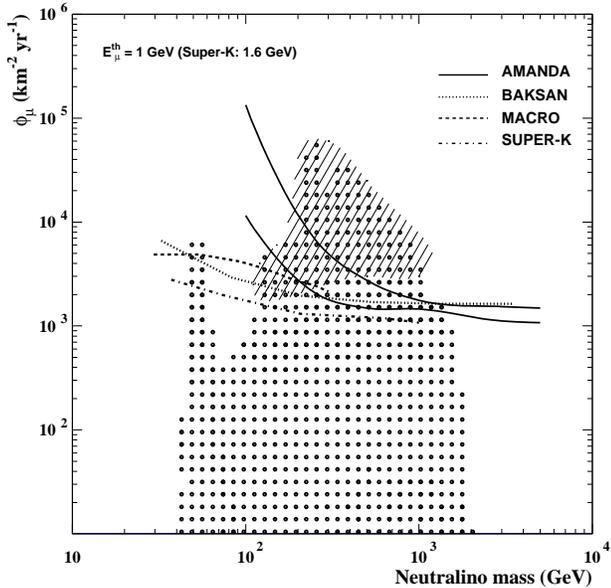,height=\linewidth,width=\linewidth}
\caption{The AMANDA limits on the muon flux from neutralino
annihilations from figure~\ref{fig:flux_limit} compared with published limits from
MACRO, Baksan and Super-Kamiokande. The dots represent model
predictions from the MSSM, calculated with the {\texttt DarkSUSY}
package~\cite{Gondolo:00a}. The dashed area shows the models disfavored by direct searches from
the DAMA collaboration~\cite{Bergstrom:98c}.}
\label{fig:limit_comparison}
\end{figure}

\section{\label{sec:comparison}Comparison with other experiments and theoretical models}

Searches for a neutrino signal from WIMP annihilation in the center
of the Earth have been performed by MACRO, Baikal, Baksan, and Super-Kamiokande. 

 In figure~\ref{fig:limit_comparison} the results 
of  Baksan~\cite{Boliev:97a}, MACRO~\cite{Ambrosio:99a} and
Super-Kamiokande~\cite{Habig:01a} are shown along with the limits from 
AMANDA obtained in the previous section and  
theoretical predictions of the MSSM as a function of WIMP mass. 
In order to be able to compare with the other experiments, the Super-Kamiokande limits have 
been scaled by a factor 1/0.9 to represent total flux limits, instead of limits based on angular cones 
including 90\% of the signal as originally presented in Ref.~\onlinecite{Habig:01a}. 
The 90\% confidence level muon flux limits for a muon energy threshold of 10~GeV published by the 
Baikal collaboration range between 
0.63$\times$10$^{4}$ km$^{-2}$~yr$^{-1}$ for a zenith half cone of 15$^\circ$ and 
0.54$\times$10$^{4}$ km$^{-2}$~yr$^{-1}$ for a zenith half cone of 5$^\circ$~(Ref. \onlinecite{Balkanov:99a}). 
Since these results are not presented as a function of 
WIMP mass, and are quoted at a slightly higher muon energy threshold, we have
not included them in the figure but we mention them here for completeness. \par

Each point in the figure represents a flux obtained with a particular 
combination of MSSM parameters, following 
Ref.~\onlinecite{Bergstrom:98b}. The sixty four original 
free parameters of the general MSSM have been reduced to seven 
by the standard assumptions about the behavior of the theory at the GUT scale
and about the supersymmetry breaking parameters in the s-fermion
sector. The independent parameters left are 
the Higgsino mass parameter $\mu$, the ratio of the Higgs vacuum 
expectation values tan$\beta$, the gaugino mass parameter
M$_{\mbox{\tiny 2}}$, the mass m$_{\mbox{\tiny A}}$ of the CP-odd Higgs boson and 
the quantities m$_{\mbox{\tiny o}}$, A$_{\mbox{\tiny t}}$ and
A$_{\mbox{\tiny b}}$ from the ansatz on  
the scale of supersymmetry breaking. These parameters were varied 
in the following ranges: -5000$\le \mu\le$5000 GeV , -5000$\le
{\mbox{M}_{\mbox{\tiny 2}}} \le$5000 GeV, 1.2$\le \mbox{tan}\beta\le$ 50, 
m$_{\mbox{\tiny A}}\le$ 1000 GeV, 100$\le \mbox{m}_{\mbox{\tiny o}}\le$ 3000
GeV, -3m$_{\mbox{\tiny o}}\le {\mbox{A}}_{\mbox{\tiny b}}\le$
3m$_{\mbox{\tiny o}}$ and 
-3m$_{\mbox{\tiny o}}\le \mbox{A}_{\mbox{\tiny t}}\le$
3m$_{\mbox{\tiny o}}$. 
Models based on parameters already excluded by accelerator limits are not shown, and the figure is 
restricted to those models which give cosmologically 
interesting neutralino relic densities, 0.025$\lesssim\Omega_{\chi}h^2<$0.5. 
A local dark matter density of 0.3 GeV/cm$^3$ has been assumed. Theoretical predictions for high mass 
neutralino models lie below the scale of the plot, since in this case
the number density of neutralinos falls down rapidly if the dark matter density is kept fixed.  \par

A complementary way to search for neutralinos is  by measuring the nuclear 
recoil in elastic neutralino-nucleus collisions on an adequate target material~\cite{Jungman:96a}.
Experiments using this direct detection technique set limits on the neutralino-nucleon cross 
section as a function of neutralino mass. The same scan over MSSM parameter space used 
to generate the theoretical points in figure~\ref{fig:limit_comparison} can be used 
to identify parameter combinations that are accessible by direct searches. 
There is not, however, a one-to-one correspondence between the results of the direct detection 
searches and the expected neutrino flux from the models probed, so comparisons with 
the results of indirect searches have to be performed with care. 
We have indicated the models disfavored by the DAMA  collaboration~\cite{Bergstrom:98c} by the 
dashed area in the figure, which has to be taken as an approximate region in view of 
the mentioned difficulties in comparing both types of detection techniques. 
We note that the models that yield high muon fluxes, and
 that are disfavored by both current results from direct searches and
by the limits shown in the figure, have in common a low value of the H$^0_2$ mass, around 92~GeV. \par

\section{\label{sec:summary}Summary}

 We have performed a search for a statistically significant excess of
vertically up-going muons with the 
AMANDA neutrino detector as a signature for neutralino annihilation
in the center of the Earth. Limits on the neutralino annihilation rate 
have been derived from the non-observation of a signal excess 
over the predicted atmospheric neutrino background. We have included the effect of the
detector systematic uncertainties and the theoretical uncertainty in
the expected number of background events in the derivation of the limits,
presenting in this way realistic limit values.\par
 A comparison with the results of MACRO, Super-K and Baksan, as well
as with theoretical expectations from the MSSM are presented.  
AMANDA, with only 130.1 days of effective exposure in 1997, has reached a sensitivity
in the high neutralino mass ($>$500 GeV) region comparable to that achieved by  
detectors with much longer live-times.\par

\begin{acknowledgments}
AMANDA is supported by the following agencies:
 The U.S. National Science Foundation, the 
University of Wisconsin Alumni Research Foundation, the  
 U.S. Department of Energy, the U.S. National Energy Research Scientific
Computing Center, the Swedish Research Council, the 
Swedish Polar Research Council, the Knut and Allice Wallenberg 
Foundation (Sweden) and the German Federal Ministry of Education and 
Research. D. F. Cowen acknowledges the support of the NSF CAREER
 program. C.P. de los Heros acknowledges support from the EU  
4$^{\mbox{\tiny th}}$ framework of Training and Mobility of Researches. P.~Loaiza was supported by the Swedish
STINT programme. We acknowledge the 
invaluable support of the Amundsen-Scott South Pole station
personnel. We are thankful to I.F.M. Albuquerque and W. Chinowsky for
 their careful reading of the manuscript and valuable comments.
\end{acknowledgments}

\bibliographystyle{natbib}

\end{document}